\documentclass{article}

\PassOptionsToPackage{numbers, compress}{natbib}

\usepackage[final]{neurips_2024_ml4ps}
\usepackage{multicol} 

\usepackage[utf8]{inputenc} 
\usepackage[T1]{fontenc}    
\usepackage{hyperref}       
\usepackage{url}            
\usepackage{booktabs}       
\usepackage{amsfonts}       
\usepackage{nicefrac}       
\usepackage{microtype}      
\usepackage{xcolor}         
\usepackage{graphicx}
\usepackage{array}

\usepackage{colortbl} 

\definecolor{lightgray}{gray}{0.9}

\usepackage{breqn}

\title{Symbolic regression for precision LHC physics}

\author{%
  Manuel Morales-Alvarado\\
  INFN, Sezione di Trieste\\
  SISSA\\
  Trieste, Italy\\
  \texttt{mmorales@sissa.it} \\
  \And
  Daniel Conde\\
  IFIC\\
  Universidad de Valencia\\
  Valencia, Spain\\
  \texttt{daniel.conde@ific.uv.es} \\
  \And
  Josh Bendavid\\
  Massachusetts Institute\\
  of Technology\\
  Cambridge, Massachusetts, USA\\
  \texttt{josh.bendavid@cern.ch} \\
  \And
  Veronica Sanz\\
  \small{IFIC}\\
  Universidad de Valencia\\
  Valencia, Spain\\
  \texttt{veronica.sanz@uv.es} \\
  \And
  Maria Ubiali\\
  \small{DAMTP}\\
  University of Cambridge\\
  Cambridge, UK\\
  \texttt{m.ubiali@damtp.cam.ac.uk} \\
}

\begin{document}

\maketitle

\begin{abstract}
We study the potential of symbolic regression (SR) to derive compact and precise analytic expressions that can improve the accuracy and simplicity of phenomenological analyses at the Large Hadron Collider (LHC). As a benchmark, we apply SR to equation recovery in quantum electrodynamics (QED), where established analytical results from quantum field theory provide a reliable framework for evaluation. This benchmark serves to validate the performance and reliability of SR before extending its application to structure functions in the Drell-Yan process mediated by virtual photons, which lack analytic representations from first principles. By combining the simplicity of analytic expressions with the predictive power of machine learning techniques, SR offers a useful tool for facilitating phenomenological analyses in high energy physics.
\end{abstract}

\section{Introduction}
\label{sec:intro}

SR is a machine learning task that discovers symbolic models by searching for simple analytic expressions that minimise both prediction error and model complexity. Unlike traditional methods, SR does not fit parameters to a potentially overparametrised model but instead finds concise formulas to describe data. This approach combines the power of machine learning with the clarity of analytical expressions, enabling the extraction of simple formulas from potentially complex datasets.

In LHC physics, some quantities have analytic expressions, while others require expensive fits or iterative algorithms for evaluation, lacking universally known formulas. Previous studies have applied SR in LHC contexts~\cite{Butter:2021rvz,Dong:2022trn}, often comparing the derived models to known analytical results. However, a key motivation for this work arises in scenarios where no reference expression exists, prompting the need to assess the reliability of SR methods.

The structure of this work is as follows. Sect.~\ref{sec:sr} briefly introduces the basics of SR. Sect.~\ref{sec:eq_rec} validates the methodology by recovering known analytical expressions from noisy data. Sect.~\ref{sec:ang_coeff} presents an SR-derived result for the Drell-Yan structure function with virtual photons, which cannot be obtained from first principles. We conclude in Sect.~\ref{sec:conc}.

\section{Symbolic regression}
\label{sec:sr}

SR is a supervised learning method that discovers closed-form analytical expressions for input-output relationships without completely predefined functional forms~\cite{Butter:2021rvz,Dong:2022trn,Choi:2010wa,Dersy:2022bym,Alnuqaydan:2022ncd}. Unlike linear regression or neural networks, SR optimises simultaneously for both accuracy and simplicity.

\begin{figure}[htp]
  \centering
  \begin{minipage}{0.30\linewidth}
    \centering
    \includegraphics[width=\linewidth]{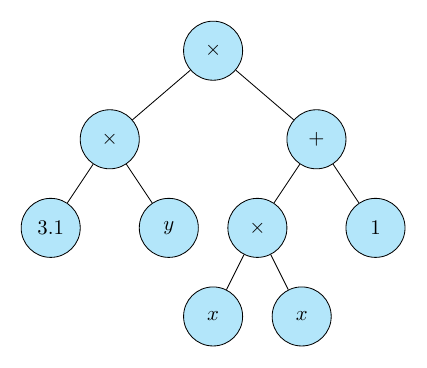} 
    \caption{Example of an expression tree.}
    \label{fig:tree}
  \end{minipage}%
  \hspace{0.05\linewidth}
  \begin{minipage}{0.30\linewidth}
    \centering
    \includegraphics[width=\linewidth]{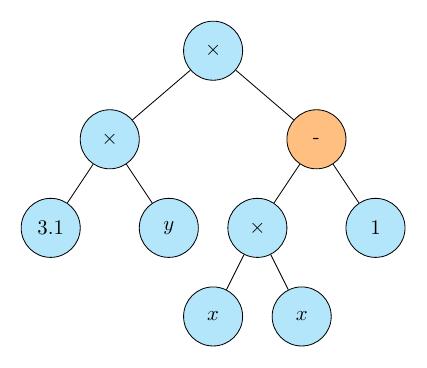} 
    \caption{Mutation of an expression tree.}
    \label{fig:tree2}
  \end{minipage}
\end{figure}

We use the PySR package~\cite{Cranmer2023InterpretableML}, a multipopulation evolutionary algorithm that evaluates symbolic expressions as expression trees. It can be highly effective for parameter spaces of moderate size~\cite{Abdussalam:2024bsm,Defranca:2023com}. Fig.~\ref{fig:tree} shows an example tree representing the equation $3.1 y \cdot (x^2 + 1)$.
In this study, equations are evaluated using the MSE loss function. During optimisation, expression trees mutate over iterations to improve based on a selection criterion, as shown in Fig.~\ref{fig:tree2}, where a '$+$' in Fig.~\ref{fig:tree} mutates to a '$-$'.

Trees can also combine through crossover, and complexity, defined by node count, is managed via multiobjective optimisation. More details are available in the original reference.

The selection criterion in PySR guides the evolutionary algorithm in choosing the fittest expression trees. There are three criteria: \texttt{accuracy}, which selects the model with the lowest loss; \texttt{score}, defined as the negative derivative of log-loss with respect to complexity, selects the model with highest decrease in loss with marginally higher complexity; and \texttt{best}, which selects the model with the highest score, provided its loss does not exceed 1.5 times that of the most accurate model.

\section{Equation rediscovery from QED}
\label{sec:eq_rec}

In this section, we apply SR to the process $e^+e^- \to \gamma^* \to \mu^+ \mu^-$ at leading order, testing its ability to recover the angular cross-section distribution in the massless limit. From QED, the distribution is:
\begin{equation}
\label{eq:eq1}
    \frac{d\sigma}{d\cos \theta} = \frac{\pi \alpha^2}{2s}(1 + \cos^2\theta),
\end{equation}
where $\theta$ is the angle between the outgoing muon and incoming electrons, $\alpha$ the QED coupling, and $s$ the squared centre-of-mass energy, treated as a constant. Radiative corrections and higher-order effects are not included, so $\alpha$ is also treated as constant. SR will aim to rediscover this equation from simulated samples.

To train the regressor, we generate $100k$ events at $\sqrt{s} = 1$ TeV using {\sc\small MadGraph5\_aMC@NLO}~\cite{Alwall:2014hca,Frederix:2018nkq} without kinematic cuts. From these events, we extract $\cos \theta$ distributions for various binnings and show the corresponding samples to the regressor. As Eq. (~\ref{eq:eq1}) indicates, $\cos \theta$ is the key kinematic variable.

In Tab.~\ref{tab:l1_results}, we present the analytical equations derived from the simulated distributions for different binnings. Comparing with Eq.~(\ref{eq:eq1}), we observe that the \texttt{accuracy} criterion often fails to recover the equation across different binning levels, fitting the noise. In contrast, \texttt{best}, successfully recovers the correct equation with finer binning. Notably, \texttt{score} consistently identifies the correct equation, even with very fine binning, demonstrating that simplicity can be an effective factor.
\begin{table}[h]
\begin{center}

\caption{Equations according to the three selection criteria for different bin sizes with $c_{\theta} \equiv \cos \theta$. The numbers that appear in these expressions have been approximated to the 5th decimal place.}
\label{tab:l1_results}

\resizebox{\textwidth}{!}{%
\begin{tabular}{@{}c>{\centering\arraybackslash}m{0.3\textwidth}>{\centering\arraybackslash}m{0.3\textwidth}>{\centering\arraybackslash}m{0.3\textwidth}@{}}
\toprule
Bins & \multicolumn{1}{c}{Accuracy} & \multicolumn{1}{c}{Best} & \multicolumn{1}{c}{Score} \\
\midrule
10 & 
$c_{\theta}(c_{\theta} + 0.00798)(0.00111 \cdot c_{\theta}^{3} + 0.03459) + 0.03503$ & 
$c_{\theta}^{2}(0.00111 \cdot c_{\theta} + 0.03459) + 0.03503$ & 
$0.03459 \cdot c_{\theta}^{2} + 0.03503$ \\
\noalign{\smallskip} \hline \noalign{\smallskip}
20 & 
$c_{\theta}(c_{\theta} + 0.01825)(-0.00155 \cdot c_{\theta}(c_{\theta} - 0.05138) + 0.03579) + 0.03485$ & 
$c_{\theta}(0.03447 \cdot c_{\theta} + 0.00064) + 0.03498$ & 
$0.03447 \cdot c_{\theta}^{2} + 0.03498$ \\
\noalign{\smallskip} \hline \noalign{\smallskip}
200 & 
$c_{\theta}^{2}(-0.64647 \cdot c_{\theta}(0.00119 \cdot c_{\theta} - 0.00151) + 0.03495) + 0.03495$ & 
$0.03447 \cdot c_{\theta}^{2} + 0.03495$ & 
$0.03447 \cdot c_{\theta}^{2} + 0.03495$ \\
\noalign{\smallskip} \hline \noalign{\smallskip}
\end{tabular}
}
\end{center}
\end{table}

Realistic simulations must account for uncertainties and fluctuations in the data. Figs.~\ref{fig:sr_pred_l1_1} and~\ref{fig:sr_pred_l1_2} show the absolute and normalised distributions for 30 bins, comparing the simulation's central value (with $1$-standard deviation Poisson uncertainty at the level of the event count in each bin), the SR result, and the analytic equation. SR demonstrates excellent agreement with the true functional form, even in cases where the simulation deviates by more than one standard deviation from it. This shows SR's ability to recover not only accurate distributions but also the correct functional dependence derived from first principles in QED.
\textbf{\begin{figure}[ht!]
  \centering
  \begin{minipage}[b]{0.49\linewidth}
    \includegraphics[width=\linewidth]{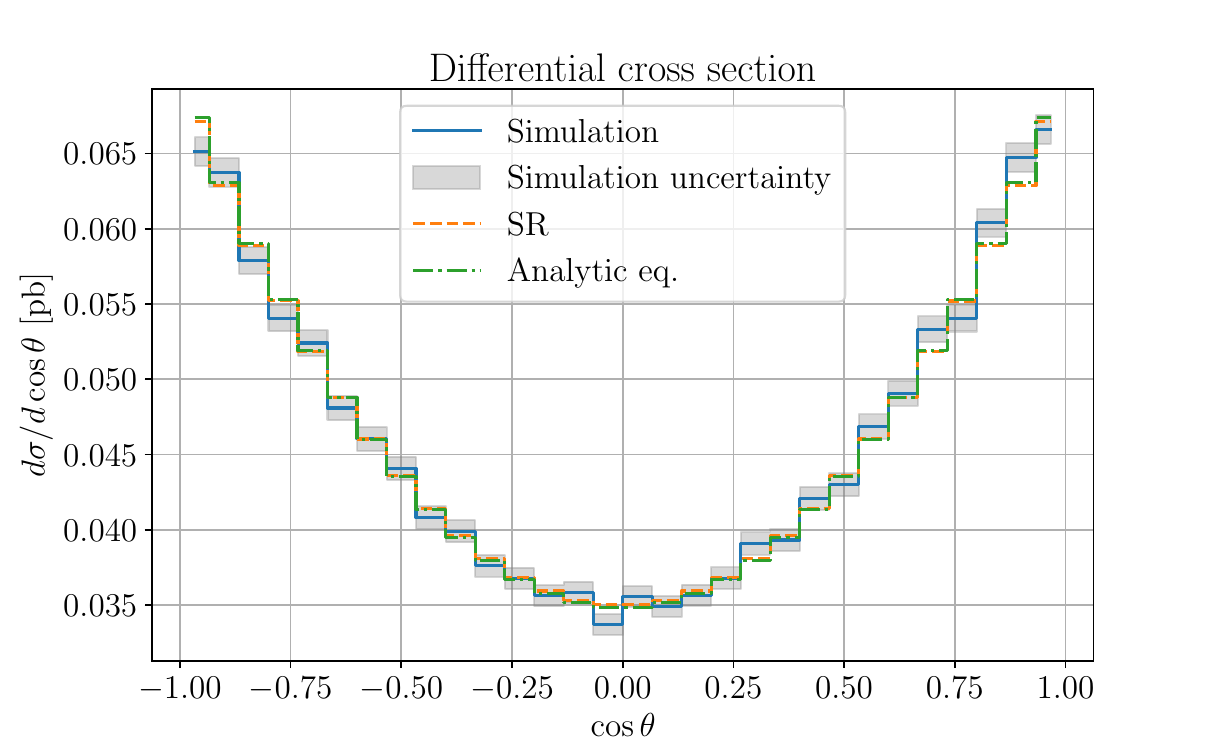}
  \caption{Absolute distribution for $30$ bins.}
  \label{fig:sr_pred_l1_1}
  \end{minipage}
  \hfill 
  \begin{minipage}[b]{0.49\linewidth}
    \includegraphics[width=\linewidth]{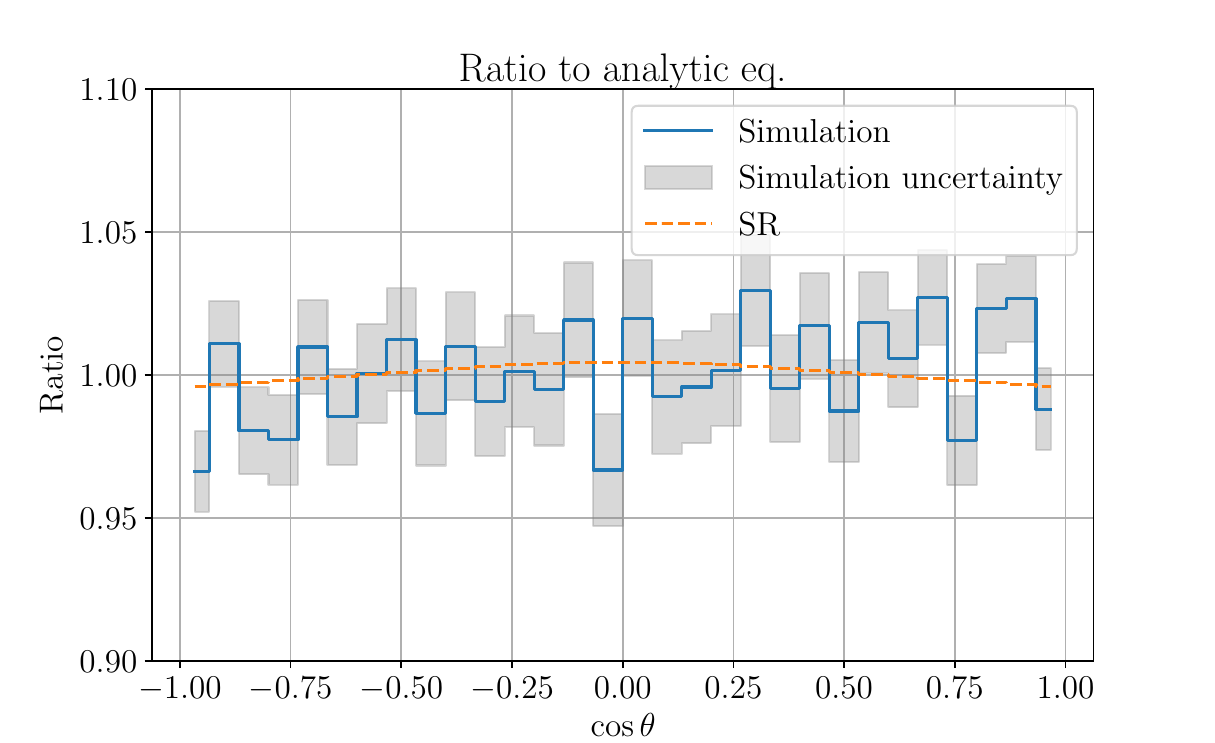}
  \caption{Normalised distribution for $30$ bins.}
  \label{fig:sr_pred_l1_2}
  \end{minipage}
\end{figure}}

Having verified a known natural law across various settings, we can now explore how SR can shed light on expressions that do not have a known closed analytical formula. This is, for example, the case of parton distribution functions and structure functions.

\section{Proton structure functions}
\label{sec:ang_coeff}

Parton distribution functions (PDFs) are essential for calculating observables at hadron colliders, as discussed in~\cite{Butterworth:2015oua, Cridge:2021qjj} and references therein. They represent momentum distribution of partons within hadrons, which must be combined with partonic cross sections to produce physical predictions to compare with experimental data. PDFs cannot be calculated from first principles, as they encapsulate the non-perturbative regime of quantum chromodynamics where the strong coupling becomes too large for perturbative methods to converge. Instead, PDFs have to be fitted from experimental data. This is achieved at the state of the art by using fixed functional forms or neural networks with hundreds of trainable weights~\cite{Bailey:2020ooq, Hou:2019efy, NNPDF:2021njg}.

Many differential observables depend on structure functions (SFs), which are weighted combinations of PDFs~\cite{Ellis:1996mzs, Peskin:1995ev}. Like PDFs themselves, SFs lack closed analytical expressions. In this section, we introduce the first SR approach to derive SFs. Our goal is to obtain accurate and compact analytical expressions that can effectively model these functions, offering a more straightforward and clear understanding of their behaviour compared to current techniques.

We consider the leading order Drell-Yan (DY) process $p \ p \to \gamma^* \to \mu^+ \ \mu^-$ at $\sqrt{{s}} = 1$ TeV and generate $100k$ events. No standard cuts are applied on the event generation. We use the CT10 NLO PDF set~\cite{Gao:2013xoa, Buckley:2014ana}. The DY double differential cross section is given by
%
\begin{equation}
\label{eq:l2_pdf}
    \frac{d^2\sigma}{dM dy} = \frac{1}{3 s} \frac{8 \pi \alpha^2}{3 M}  \left( 
\sum_{q} Q_{q}^2 (f_{q}(x_1, \tau) f_{\overline{q}} (x_2, \tau) + f_{\overline{q}}(x_1, \tau) f_{q}(x_2, \tau) \right) \equiv \frac{1}{3 s} \frac{8 \pi \alpha^2}{3 M} F(M, y),
\end{equation}
where $M$ and $y$ are, respectively, the invariant mass and the rapidity of the muon pair, $x_1 = \sqrt{\tau} e^y$ and $x_2 = \sqrt{\tau} e^{-y}$ are the parton momentum fractions that cannot be bigger than $1$, and $\tau = M^2 / s$. The sum in parentheses runs over the $q=u, d, s, c$ quark flavours, $Q_{q}$ is their respective electric charge, and $f_{q}$ are their associated PDFs. We are exclusively interested in the SF $F(M, y)$ at high resolution in the kinematic coverage as the rest of the distribution is known. 

\begin{figure}[ht!]
  \centering
  \begin{minipage}[b]{0.49\linewidth}
    \includegraphics[width=\linewidth]{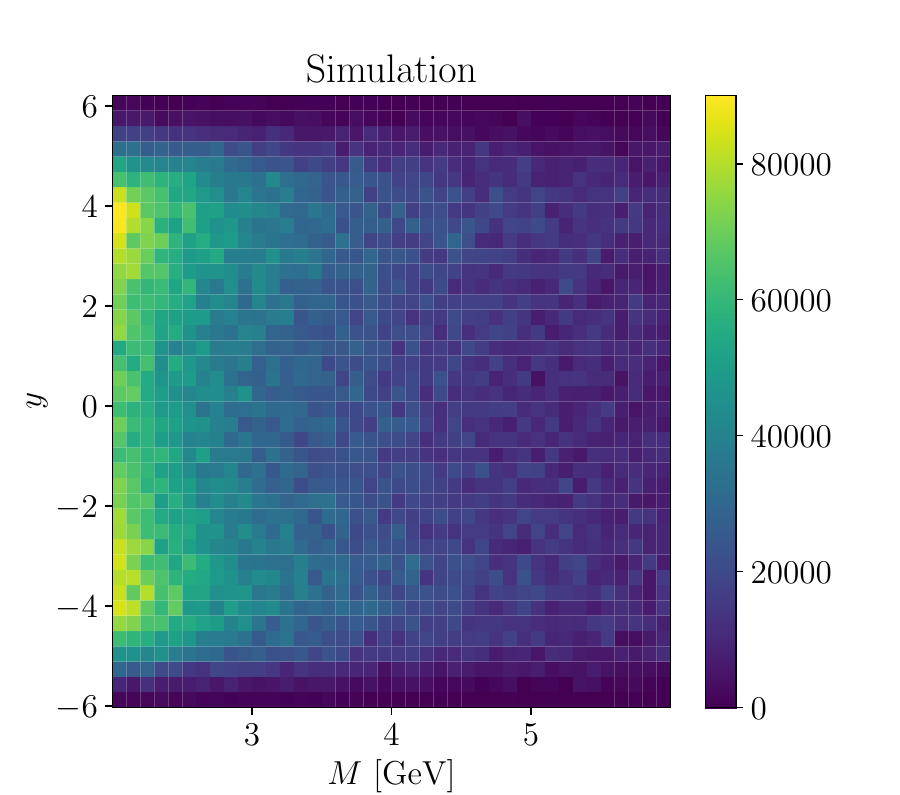}
    \caption{SF values obtained from the reweighted simulation.}
    \label{fig:l2_mg}
  \end{minipage}
  \hfill 
  \begin{minipage}[b]{0.49\linewidth}
    \includegraphics[width=\linewidth]{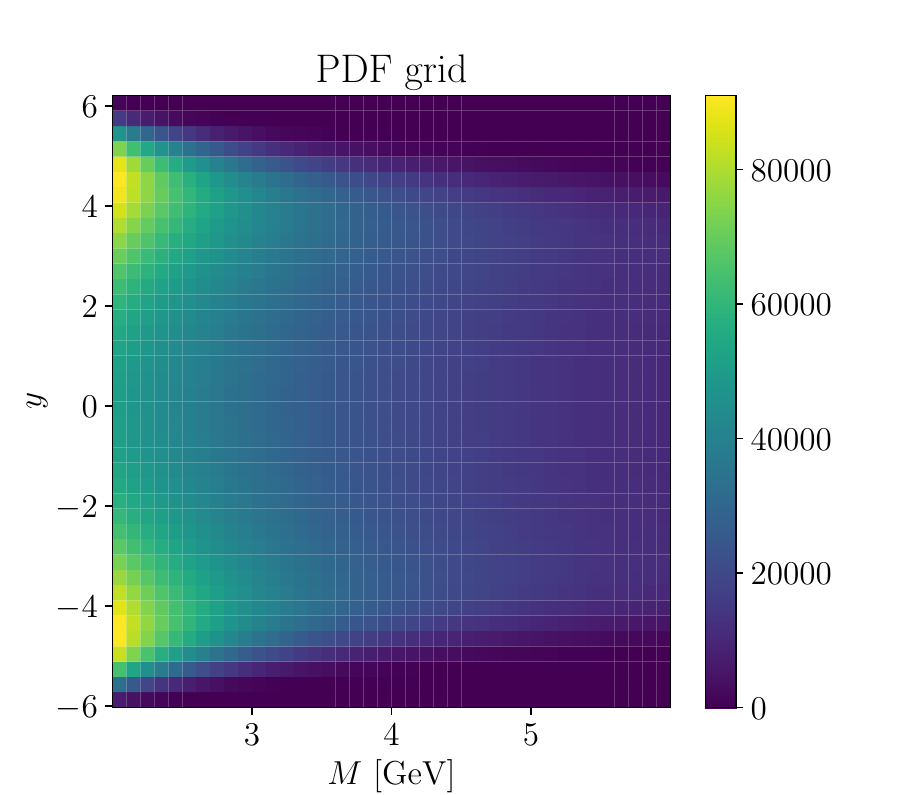}
    \caption{SF values obtained from the central value PDF grids.}
    \label{fig:l2_sr}
  \end{minipage}
  
  
  \begin{minipage}[b]{0.5\linewidth}
    \centering 
    \includegraphics[width=\linewidth]{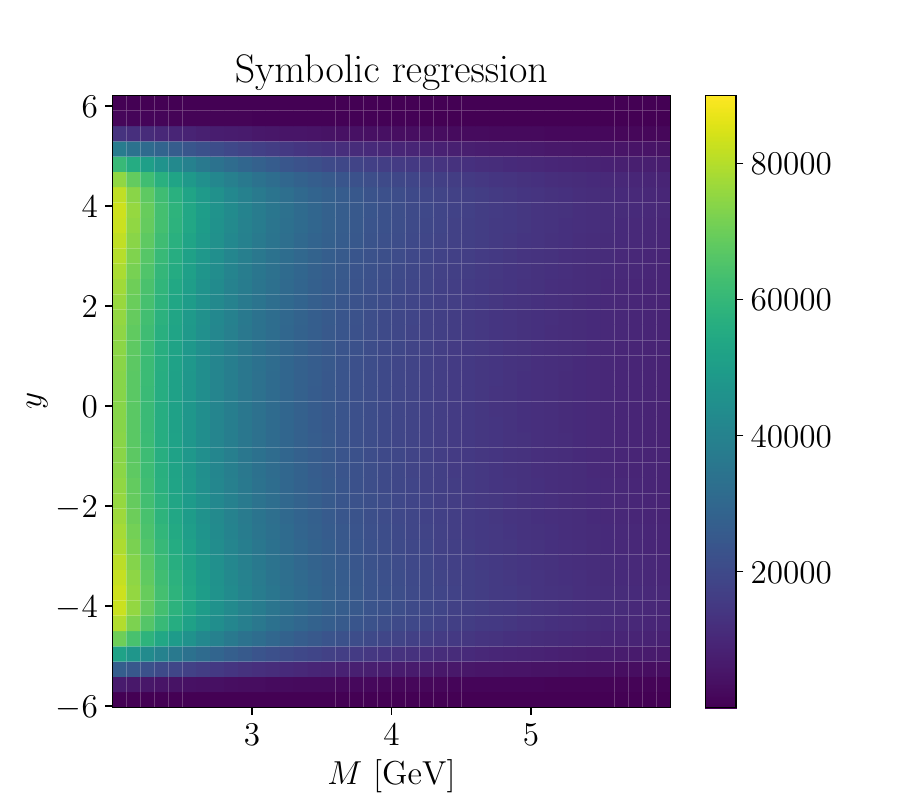}
    \caption{SF $F(M, y)$ values obtained with SR.}
    \label{fig:l2_lhapdf}
  \end{minipage}
\end{figure}

The SF values from the simulation are shown in Fig.~\ref{fig:l2_mg}, calculated by reweighting the double differential distribution with the prefactor $ \frac{1}{3s} \frac{8 \pi \alpha^2}{3M} $ as per Eq.~(\ref{eq:l2_pdf}). Applying SR to this distribution yields a hall-of-fame set of models at various complexities, with selected examples shown in Table~\ref{tab:selected_equations}. The \texttt{best} model, with a complexity of $33$ (highlighted in gray), is shown in Fig.~\ref{fig:l2_sr}. These results are compared to the central replica values of the PDF set, shown in Fig.~\ref{fig:l2_lhapdf}.
From the figures, we observe that the SR result provides a smooth function that approximates well the PDF grid. It successfully extrapolates to the unphysical regions of the kinematic coverage by suppressing the function (although it does not fully reach zero) and effectively reduces fluctuations at high invariant mass $M$. Moreover, the expressions in the table show that the SF can be parametrised at low complexity ($3$) using inverse power laws in $M$. At higher complexities, where there is resolution in $y$, the SR recognises that it must be symmetric in $y$, as dictated by particle kinematics. Notably, the most complex expression (complexity $35$), while more accurate by construction, achieves a lower score due to increased expression length.
\begin{table}[h]
\begin{center}
\caption{Selection of \texttt{best} SR expressions $F(M, y)$ with their complexities and scores. Constants are approximated for display purposes.}
\label{tab:selected_equations}

\resizebox{\textwidth}{!}{%
\begin{tabular}{@{}c>{\centering\arraybackslash}m{0.5\textwidth}>{\centering\arraybackslash}m{0.2\textwidth}@{}}
\toprule
Complexity & \multicolumn{1}{c}{Equation} & \multicolumn{1}{c}{Score} \\
\midrule
$3$ & $\frac{9.16 \cdot 10^{4}}{M}$ & $0.359$\\

\rowcolor{lightgray} 
$33$ & 
\begin{minipage}{0.5\textwidth} 
\vspace{-0.7em}
\[
\frac{2.86 \cdot 10^{5} \left(0.0461 \cdot 1.15^{y^2}\right)^{- 0.0250 \cdot 1.15^{y^2}}}{M^{2}}
\]
\end{minipage} 
& $0.387$ \\
$35$ & 
\begin{minipage}{0.5\textwidth} 
\vspace{-0.5em}
\[
\frac{2.86 \cdot 10^{5} \left(0.0461 \cdot 1.15^{y^2}\right)^{- 0.0250 \cdot 1.15^{y^2}}}{M^{2} + 0.117}
\]
\end{minipage} 
& $0.00967$ \\
\bottomrule
\end{tabular}
}
\end{center}
\end{table}

To compare the simulation and SR results to the PDF grid baseline, in Table~\ref{tab:metrics_comparison} we use the following fit quality metrics:
\begin{multicols}{3}
\noindent
\[
\text{RMSE} = \sqrt{\frac{1}{n} \sum_{i=1}^{n} (y_i - \hat{y}_i)^2},
\]
\columnbreak
\[
\text{MAE} = \frac{1}{n} \sum_{i=1}^{n} |y_i - \hat{y}_i|,
\]
\columnbreak
\[
R^2 = 1 - \frac{\sum_{i=1}^{n} (y_i - \hat{y}_i)^2}{\sum_{i=1}^{n} (y_i - \bar{y})^2},
\]
\end{multicols}
\vspace{-1.5em}
where \( n \) is the number of data points, \( y_i \) the actual value, \( \hat{y}_i \) the predicted value, and \( \bar{y} \) the mean of \( y_i \). We see that SR achieves lower RMSE and MAE values, indicating a better overall fit quality with the PDF grid. Additionally, the higher $R^2$ value for the SR fit ($0.9030$ vs. $0.8898$) reflects a stronger correlation. These results demonstrate the capability of SR to model the data with good precision.
%
\begin{table}[h]
\begin{center}
\caption{Comparison of metrics of the reweighted simulation and the SR fit with respect to the PDF grid.}
\label{tab:metrics_comparison}

\resizebox{\textwidth}{!}{%
\begin{tabular}{@{}c>{\centering\arraybackslash}m{0.25\textwidth}>{\centering\arraybackslash}m{0.25\textwidth}@{}}
\toprule
Metric & \multicolumn{1}{c}{Reweighted simulation} & \multicolumn{1}{c}{Symbolic regression} \\
\midrule
Root mean square error (RMSE) & $6.09 \times 10^{3}$ & $5.72 \times 10^{3}$ \\
Mean absolute error (MAE) & $4.26 \times 10^{3}$ & $3.74 \times 10^{3}$ \\
Coef. of determination ($R^2$) & $0.8898$ & $0.9030$ \\
\bottomrule
\end{tabular}
}
\end{center}
\end{table}
%
\vspace{-1.0em}

\section{Conclusion}
\label{sec:conc}

We have explored the application of SR to derive simple yet accurate analytical formulas in the context of collider phenomenology. Using well-established QED processes as a benchmark, we validated the reliability of SR by recovering known analytical expressions from noisy data under varying conditions. Furthermore, we extended this methodology to SFs in Drell-Yan, showing the potential of SR to provide accurate and simple results even in scenarios where no closed-form solutions are available. This study highlights the utility of SR in simplifying and enhancing the analysis of complex datasets in high energy physics, and could be extended to study more sophisticated uncertainties, higher-dimensional distributions, or higher-order processes like those involving electroweak boson production with angular coefficients~\cite{Collins:1977iv, Mirkes:1992hu, Mirkes:1994dp, Mirkes:1994eb, Mirkes:1994nr, ATLAS:2023lsr}. 

SR combines machine learning with analytical expressions, enabling accurate closed-form models from complex datasets. This work contributes to precision physics at the LHC and, more in general, machine learning-assisted discovery in high energy physics.

\newpage
\bibliographystyle{unsrt}  
\bibliography{references}  

\begin{thebibliography}{10}

\bibitem{Butter:2021rvz}
Anja Butter, Tilman Plehn, Nathalie Soybelman, and Johann Brehmer.
\newblock {Back to the Formula -- LHC Edition}.
\newblock 9 2021.

\bibitem{Dong:2022trn}
Zhongtian Dong, Kyoungchul Kong, Konstantin~T. Matchev, and Katia Matcheva.
\newblock {Is the machine smarter than the theorist: Deriving formulas for
  particle kinematics with symbolic regression}.
\newblock {\em Phys. Rev. D}, 107(5):055018, 2023.

\bibitem{Choi:2010wa}
Suyong Choi.
\newblock {Construction of a Kinematic Variable Sensitive to the Mass of the
  Standard Model Higgs Boson in $H -> WW^* -> l^+ \nu l^- \bar{\nu}$ using
  Symbolic Regression}.
\newblock {\em JHEP}, 08:110, 2011.

\bibitem{Dersy:2022bym}
Aur\'elien Dersy, Matthew~D. Schwartz, and Xiaoyuan Zhang.
\newblock {Simplifying Polylogarithms with Machine Learning}.
\newblock 6 2022.

\bibitem{Alnuqaydan:2022ncd}
Abdulhakim Alnuqaydan, Sergei Gleyzer, and Harrison Prosper.
\newblock {SYMBA: symbolic computation of squared amplitudes in high energy
  physics with machine learning}.
\newblock {\em Mach. Learn. Sci. Tech.}, 4(1):015007, 2023.

\bibitem{Cranmer2023InterpretableML}
M.~Cranmer.
\newblock Interpretable machine learning for science with pysr and
  symbolicregression.jl.
\newblock {\em ArXiv}, abs/2305.01582, 2023.

\bibitem{Abdussalam:2024bsm}
Shehu AbdusSalam, Steve Abel, and Miguel~Crispim Romao.
\newblock Symbolic regression for beyond the standard model physics.
\newblock 2024.

\bibitem{Defranca:2023com}
F.~O. de~Franca, M.~Virgolin, M.~Kommenda, M.~S. Majumder, M.~Cranmer,
  G.~Espada, L.~Ingelse, A.~Fonseca, M.~Landajuela, B.~Petersen, R.~Glatt,
  N.~Mundhenk, C.~S. Lee, J.~D. Hochhalter, D.~L. Randall, P.~Kamienny,
  H.~Zhang, G.~Dick, A.~Simon, B.~Burlacu, Jaan Kasak, Meera Machado, Casper
  Wilstrup, and W.~G.~La Cava.
\newblock Interpretable symbolic regression for data science: Analysis of the
  2022 competition.
\newblock 2023.

\bibitem{Alwall:2014hca}
J.~Alwall, R.~Frederix, S.~Frixione, V.~Hirschi, F.~Maltoni, O.~Mattelaer,
  H.~S. Shao, T.~Stelzer, P.~Torrielli, and M.~Zaro.
\newblock {The automated computation of tree-level and next-to-leading order
  differential cross sections, and their matching to parton shower
  simulations}.
\newblock {\em JHEP}, 07:079, 2014.

\bibitem{Frederix:2018nkq}
R.~Frederix, S.~Frixione, V.~Hirschi, D.~Pagani, H.~S. Shao, and M.~Zaro.
\newblock {The automation of next-to-leading order electroweak calculations}.
\newblock {\em JHEP}, 07:185, 2018.
\newblock [Erratum: JHEP 11, 085 (2021)].

\bibitem{Butterworth:2015oua}
Jon Butterworth et~al.
\newblock {PDF4LHC recommendations for LHC Run II}.
\newblock {\em J. Phys. G}, 43:023001, 2016.

\bibitem{Cridge:2021qjj}
Thomas Cridge.
\newblock {PDF4LHC21: Update on the benchmarking of the CT, MSHT and NNPDF
  global PDF fits}.
\newblock {\em SciPost Phys. Proc.}, 8:101, 2022.

\bibitem{Bailey:2020ooq}
S.~Bailey, T.~Cridge, L.~A. Harland-Lang, A.~D. Martin, and R.~S. Thorne.
\newblock {Parton distributions from LHC, HERA, Tevatron and fixed target data:
  MSHT20 PDFs}.
\newblock {\em Eur. Phys. J. C}, 81(4):341, 2021.

\bibitem{Hou:2019efy}
Tie-Jiun Hou et~al.
\newblock {New CTEQ global analysis of quantum chromodynamics with
  high-precision data from the LHC}.
\newblock {\em Phys. Rev. D}, 103(1):014013, 2021.

\bibitem{NNPDF:2021njg}
Richard~D. Ball et~al.
\newblock {The path to proton structure at 1\% accuracy}.
\newblock {\em Eur. Phys. J. C}, 82(5):428, 2022.

\bibitem{Ellis:1996mzs}
R.~Keith Ellis, W.~James Stirling, and B.~R. Webber.
\newblock {\em {QCD and collider physics}}, volume~8.
\newblock Cambridge University Press, 2 2011.

\bibitem{Peskin:1995ev}
Michael~E. Peskin and Daniel~V. Schroeder.
\newblock {\em {An Introduction to quantum field theory}}.
\newblock Addison-Wesley, Reading, USA, 1995.

\bibitem{Gao:2013xoa}
Jun Gao, Marco Guzzi, Joey Huston, Hung-Liang Lai, Zhao Li, Pavel Nadolsky, Jon
  Pumplin, Daniel Stump, and C.~P. Yuan.
\newblock {CT10 next-to-next-to-leading order global analysis of QCD}.
\newblock {\em Phys. Rev. D}, 89(3):033009, 2014.

\bibitem{Buckley:2014ana}
Andy Buckley, James Ferrando, Stephen Lloyd, Karl Nordstr\"om, Ben Page, Martin
  R\"ufenacht, Marek Sch\"onherr, and Graeme Watt.
\newblock {LHAPDF6: parton density access in the LHC precision era}.
\newblock {\em Eur. Phys. J. C}, 75:132, 2015.

\bibitem{Collins:1977iv}
John~C. Collins and Davison~E. Soper.
\newblock {Angular Distribution of Dileptons in High-Energy Hadron Collisions}.
\newblock {\em Phys. Rev. D}, 16:2219, 1977.

\bibitem{Mirkes:1992hu}
E.~Mirkes.
\newblock {Angular decay distribution of leptons from W bosons at NLO in
  hadronic collisions}.
\newblock {\em Nucl. Phys. B}, 387:3--85, 1992.

\bibitem{Mirkes:1994dp}
E.~Mirkes and J.~Ohnemus.
\newblock {Angular distributions of Drell-Yan lepton pairs at the Tevatron:
  Order $\alpha-s^{2}$ corrections and Monte Carlo studies}.
\newblock {\em Phys. Rev. D}, 51:4891--4904, 1995.

\bibitem{Mirkes:1994eb}
E.~Mirkes and J.~Ohnemus.
\newblock {$W$ and $Z$ polarization effects in hadronic collisions}.
\newblock {\em Phys. Rev. D}, 50:5692--5703, 1994.

\bibitem{Mirkes:1994nr}
E.~Mirkes and J.~Ohnemus.
\newblock {Polarization effects in Drell-Yan type processes h1 + h2
  ---\ensuremath{>} (W, Z, gamma*, J / psi) + x}.
\newblock In {\em {1994 Meeting of the American Physical Society, Division of
  Particles and Fields (DPF 94)}}, pages 1721--1723, 8 1994.

\bibitem{ATLAS:2023lsr}
Georges Aad et~al.
\newblock {A precise measurement of the Z-boson double-differential transverse
  momentum and rapidity distributions in the full phase space of the decay
  leptons with the ATLAS experiment at $\sqrt s$ = 8 TeV}.
\newblock 9 2023.

\end{thebibliography}


\end{document}